\documentclass[aps,prb,groupedaddress,floatfix]{revtex4}

\usepackage{graphicx}
\usepackage{subfigure}
\usepackage{pstricks,pst-node,pst-text,pst-3d,graphpap,pst-plot}
\def\be{\begin{equation}}
\def\ee{\end{equation}}
\def\bea{\begin{eqnarray}}
\def\eea{\end{eqnarray}}

\begin{document}
\def\gC{\mbox{\boldmath $C$}}
\def\gZ{\mbox{\boldmath $Z$}}
\def\gR{\mbox{\boldmath $R$}}
\def\gN{\mbox{\boldmath $N$}}
\def\ua{\uparrow}
\def\da{\downarrow}
\def\a{\alpha}
\def\b{\beta}
\def\g{\gamma}
\def\G{\Gamma}
\def\d{\delta}
\def\D{\Delta}
\def\e{\epsilon}
\def\ve{\varepsilon}
\def\z{\zeta}
\def\h{\eta}
\def\th{\theta}
\def\k{\kappa}
\def\l{\lambda}
\def\L{\Lambda}
\def\m{\mu}
\def\n{\nu}
\def\x{\xi}
\def\X{\Xi}
\def\p{\pi}
\def\P{\Pi}
\def\r{\rho}
\def\s{\sigma}
\def\S{\Sigma}
\def\t{\tau}
\def\f{\phi}
\def\vf{\varphi}
\def\F{\Phi}
\def\c{\chi}
\def\w{\omega}
\def\W{\Omega}
\def\Q{\Psi}
\def\q{\psi}
\def\de{\partial}
\def\inf{\infty}
\def\ra{\rightarrow}
\def\bra{\langle}
\def\ket{\rangle}
\newcommand{\rigalunga}{\rule{\textwidth}{1pt}}

\title{Correlated Nanoscopic  Josephson Junctions }


\author{Stefano Bellucci$^1$ Michele Cini$^{1,2}$, Pasquale Onorato$^{1,3}$, and Enrico Perfetto$^{1,4}$ \\}
\address{
       $^1$ INFN, Laboratori Nazionali di Frascati,
        P.O. Box 13, 00044 Frascati, Italy. \\
        $^2$ Dipartimento di Fisica,
        Universit\`{a} di Roma Tor Vergata, Via della Ricerca Scientifica, Roma, Italy\\
       $^3$ Dipartimento di Scienze Fisiche,
        Universit\`{a} di Roma Tre, Via della Vasca Navale 84, 00146 Roma, Italy\\
        $^4$ Instituto de Estructura de la Materia.
        Consejo Superior de Investigaciones Cient{\'\i}ficas.
        Serrano 123, 28006 Madrid. Spain}
\date{\today}

\begin{abstract}
 We discuss  correlated lattice models with a time-dependent
 potential  across a barrier and show how to implement a
  Josephson-junction-like behavior. The pairing occurs
  by a correlation effect enhanced by the symmetry of the system. In order to produce the effect  we
  need
   a mild distortion  which causes  avoided
crossings in the many-body spectrum. The Josephson-like response
involves
  a quasi-adiabatic evolution in the time-dependent field.  Besides,
  we observe an  inverse-Josephson (Shapiro) current  by applying an AC bias;
   a supercurrent in the absence of electromotive force can also be
excited. The qualitative arguments are supported by explicit exact
solutions in prototype 5-atom clusters with  on-site repulsion.
These basic units are then combined in ring-shaped  systems, where
one of the units sits at a higher potential and works as a barrier.
In this case the solution is found by mapping the low-energy
Hamiltonian into  an effective anisotropic  Heisenberg chain. Once again,
we present evidence for a superconducting flux quantization, i.e. a
Josephson-junction-like behavior suggesting the build-up of an
effective order parameter already in few-electron systems. Some
general implications for the quantum theory of transport are also
briefly discussed, stressing the nontrivial occurrence of asymptotic
current oscillations for long times in the presence of bound states.

\end{abstract}

\maketitle

\section{Introduction}

The Josephson effect is one of the clearest fingerprints from which
one can diagnose superconductivity. A DC   electromotive force
(e.m.f.) $V$  across a thin insulating layer in a
superconductor-insulator-superconductor junction
produces\cite{josephson} an AC current response. The Josephson
\index{Josephson }current  is given by the well known
expression\cite{grosso}
 \be\label{Josephson} I=I_{J}\sin\left(
\frac{2e V}{\hbar}t+\g_{0}\right),\ee where $I_{J}$ and $\g_{0}$ are
 constants. In textbooks this result is explained in terms of the
\emph{order parameter} $\psi$ at the junction; within the insulator
one has \be \psi(z)=\psi_{left}e^{-\a z}+\psi_{right}e^{\a
(z-s)},\ee
 where $s$
is the barrier width,   $\a$ is a characteristic  inverse length and
$z$ is the normal coordinate; $\psi_{left}$ and $\psi_{right}$ are
complex constants. The quantum mechanical current $I$ is computed by
the Ginzburg-Landau formula, that is, by using $\psi$ as a
single-particle wave function. Thus, $I$ is proportional to
$\sin\theta$, where $\theta$ is the phase difference between
$\psi_{left}$ and $\psi_{right}$. Letting $\theta(t)=\frac{2e
V}{\hbar}t+\g_{0}$ we obtain (\ref{Josephson}).

The electrons cannot come into play in this picture, since the order
parameter is a macroscopic concept.  A very enlightening model
argument which is microscopic in spirit is the one due to Ferrel and
Prange\cite{ferrel}. Let $H_{T}$ denote the electron tunneling
Hamiltonian between the left and right superconductors. The junction
with $N-\n$ pairs on the left and $\n$ on the right (both numbers
being huge) is mapped to a tight-binding chain described by an
effective Hamiltonian $H_{F}$ that describes pair hopping. For
$H_{T}\rightarrow 0$ the energy $E_{0}$ of the junction does not
depend on $\n$. So, the state is defined by the amplitude
$\phi_{\n}$ obtained from

\be H_{F}\phi_{\n}=E_{0} \phi_{\n} +F(\phi_{\n+1}+\phi_{\n-1}),\ee
 where the pair hopping matrix element arising to the second-order
in $H_{T}$
 reads

\be F=\sum_{i}\frac{\langle \n+1|H_{T}|i\rangle\langle i
 | H_{T}|\n\rangle}{E_{0}-E_{i}}\ee
 and the sum is  over the intermediate
states $i$ of energy $E_{i}$ with an electron in the barrier. The
eigenfunctions of $H_{F}$ are plane waves in $\n$ space, with
eigenvalues
 \be \e(k)=E_{0}+2F\cos
k.\ee A current
\begin{equation}\label{ferrel} I=\frac{4 e F}{\hbar}\sin k
\end{equation} is associated to the group velocity $\frac{1}{\hbar}\frac{\de
E}{\de k}$ in the fictitious chain and to the real current across
the unbiased junction. A wave packet centered at $\bar{k}$ in the
fictitious chain would be accelerated by \be \frac{d}{dt}  \hbar
\bar{k}
  = 2 e V\ee
 and this inserted into
(\ref{ferrel})\index{ferrel} reproduces (\ref{Josephson}) again,
seen from another angle. The description by Ferrel and Prange is
still aimed at a macroscopic system with huge pair numbers enabling
us to use the concept of a group velocity.

The question that arises is: what happens in small systems, when the
fictitious  chain is  very short and the   group velocity has no
meaning? The Ferrel - Prange description must be modified, however
pairs will continue to hop as bound units, and the AC response to
a DC bias is a qualitative feature which looks well suited to test
any model for superconductivity. Being a qualitative feature, it is
hard to see how it can depend on size. Moreover, the order parameter
which is well defined only in the thermodynamic limit must be
supplemented by a microscopic description valid in finite systems;
there, such concepts like long-range order do not apply but clear
signatures like lack of dissipation must still distinguish
superconducting from normal currents. The very nature of these
signatures deserves investigation.

 In this paper we wish to look for   precursors to the Josephson behavior
far from the thermodynamic limit; small cluster studies are
motivated by conceptual as well as practical reasons.
 Indeed, the growing technological   interest in nano-systems justifies the
identification and search. We focus on repulsive Hubbard-like models
that we want to test for correlation-induced superconductivity; this
is relevant to the search for possible non-conventional mechanisms,
that have been considered by several workers e.g. in the context of
high-T$_{C}$  materials. The choice of the geometry has been
prompted by previous studies\cite{topicalreview} in the framework of
the $W=0$ theory; this develops  the role of symmetry in inducing
pairing from repulsive interactions. Besides, we also found the
conditions leading to superconducting flux quantization in
mesoscopic systems. We are therefore interested in testing this
mechanism in time-dependent conditions.

Below we use a Gedanken experiment for understanding the problem.
Developing this idea, one meets the difficulty that small systems
always give AC response to a DC bias for quantum mechanical
reasons; that is, one finds oscillating currents analogous to Rabi
oscillations. However, the frequency of the oscillations is
proportional to the charge, so a sharp effect of pairing remains,
i.e. supercurrents are signaled by doubled frequencies. This is a
conceptual problem and we shall be in position to  add more to try
to clarify this point in the conclusions.

 There are further reasons of interest in this kind of
problems. Transient current response are now actively studied, and
within the TDLDA the asymptotic behavior of currents yields the
static \index{Stefanucci} current-voltage
characteristic\cite{stefanucci04}. Here we shall find a class of
systems where, owing to correlation induced bound states, the asymptotic
behavior is {\em oscillatory.} There is a clear need to develop new
techniques, capable of tackling such situations.

Here we outline the plan of this paper. In Section \ref{w0} we summarize
the theory of $W=0$ pairing in Hubbard clusters. Section
\ref{unosolo} presents our   one-unit model.  Here we discuss the
simplest case, namely a 5-site Hubbard cluster with square symmetry.
For short, we shall refer to it as the  CuO$_{4}$ unit; this is
known to yield pairing and flux quantization, and here we
show that it is also a prototype system for dynamic phenomena. One
must  perturb the square symmetry, in order to mimic a barrier in
such a tiny system. The current excited by a constant electro-moving
force (emf) is oscillatory with a frequency which is exactly the
same as the Josephson frequency in Eq. (\ref{Josephson}); moreover
we shall  observe Shapiro spikes (inverse Josephson effect)  which
represent a DC response to an AC bias at certain frequencies. A
constant current (with some periodic ripple) flows in the absence of
an applied potential difference, which is typical of superconducting
systems. All these effects disappear when pairing fails, e.g. if the
repulsion $U$ is removed. In Section \ref{many}   we shall extend
the investigation of Josephson-like  currents to large systems that
are rings composed of any number of Cu$O_{4}$ units, up to the
thermodynamic limit in principle.  For small inter-unit hopping, the
many-electron system can be exactly mapped into a
Heisenberg-Ising-like chain. Systems hosting many pairs are
described. An effective barrier is simulated by lifting the one-body
levels of one of the Cu$O_{4}$ clusters; the superconducting flux
quantization is again obtained and time-dependent solutions with a
constant emf are also presented. Finally, we present our conclusions
in Section \ref{Conclusion}.

\section{W=0 pairing}\label{w0}

In the context of High-T$_{C}$ superconductivity, several approaches
based on the  Hubbard Hamiltonian  have appeared in the literature.
The quantity \be \D(N)=E(N)+E(N-2)-2E(N-1), \ee
 where $E(N)$ is the
ground state energy of the system with $N$ fermions is used to
evaluate the effective interaction between particles in the
interacting ground state.  A negative   $\D$ means that an effective
attraction is developing from the repulsive interactions. We refer
to Ref. \onlinecite{topicalreview} for the general theory based on
the fact that highly-symmetric clusters possess 2-particle singlet
eigenstates which do not feel the on-site repulsion $W$; these are
called      $W=0$ pairs. The persistence of the ``$W=0$'' character
in the interacting case is discussed in detail there, with the flux
quantization properties and their group theory analysis. The
CuO$_{4}$ cluster  with Hamiltonian  \be
 H_{0}= \e_{p}\sum_{\s,i=2}^{5}p_{i,\s}^{\dag}p_{i,\s}+ t_{pd} \sum_{i\sigma}(
d^{\dagger}_{\alpha \sigma}p_{\alpha, i\sigma}+ p_{\alpha,
i\sigma}^{\dagger}d_{\alpha \sigma})+ U(n^{(d)}_{\alpha \uparrow}
n^{(d)}_{\alpha \downarrow}+\sum_{i}n^{(p)}_{\alpha, i
\uparrow}n^{(p)}_{\alpha, i\downarrow}), \ee
 is the simplest one to give the effect; here,
$p^{\dagger}_{\sigma, i}$ is the  creation operator for a fermion
onto the Oxygen $i=2,..,5,$  $d$ destroys a fermion on the central
site 1; besides, $\e_{p}$ is the O energy level, $t_{pd}$ is the p-d
hopping matrix element, and $U$ is the Hubbard repulsion. In a wide
range of parameters, this model yields $\D<0$ for $N=4.$

 The magnetic properties of the CuO$_{4}$ cluster will be
discussed below in more detail.


\vspace{1cm}

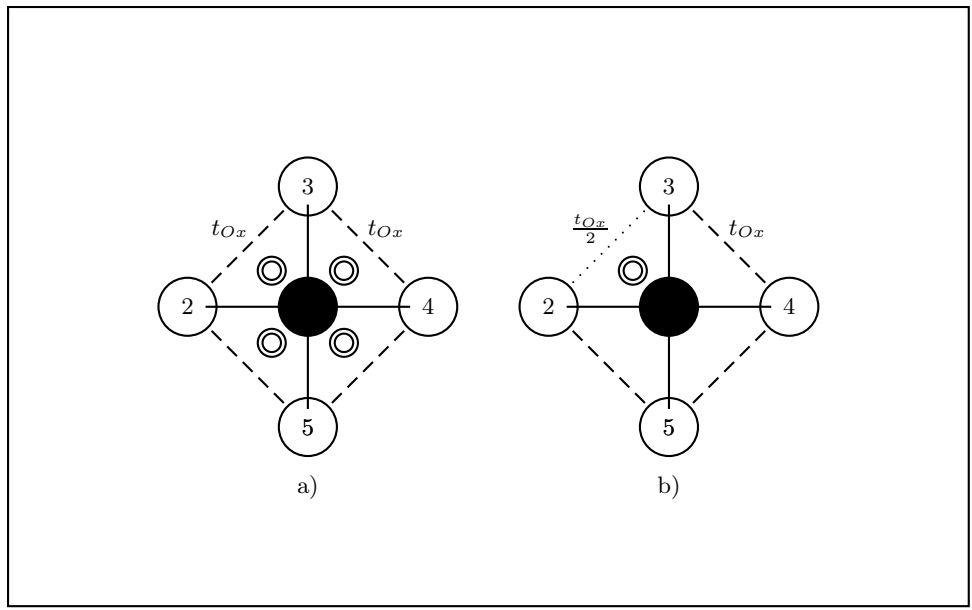
\begin{figure}\label{unit}
\begin{center}\psset{xunit=.8cm,yunit=.8cm}
 \begin{pspicture*}(-8,-30)(8,-20) \psframe(-8,-30)(8,-20)
 \rput(0,-20){
 \rput(-3,-5){
 \pscircle(0,-2){.4}\rput(0,-2){5}\pscircle(0,2){.4}
\pscircle(-2,0){.4}\pscircle(2,0){.4}
  \pscircle*[linewidth=.07](0,0){.4}
  \psline[](0,0)(0,1.7) \psline[](0,0)(1.7,0) \psline[](0,0)(-1.7,0)
   \psline[](0,0)(0,-1.7)
   \psline[linestyle=dashed](1.6,0.4)(0.4,1.6)
   \psline[linestyle=dashed](-1.6,-.4)(-0.4,-1.6)

      \psline[linestyle=dashed](1.6,-0.4)(.4,-1.6)
   \psline[linestyle=dashed](-0.4,1.6)(-1.6,.4)
   \rput(0,-2){5}\rput(0,2){3}\rput(-2,0){2}\rput(2,0){4}
\rput(1.3,1.3){$t_{Ox}$}\rput(-1.3,1.3){$t_{Ox}$}
\pscircle[doubleline=true](-.6,.6){.2}
\pscircle[doubleline=true](.6,.6){.2}
\pscircle[doubleline=true](-.6,-.6){.2}
\pscircle[doubleline=true](.6,-.6){.2} \rput(0,-3){a)}
 }
\rput(3,-5){\pscircle[doubleline=true](-.6,.6){.2}
 \pscircle(0,-2){.4}\rput(0,-2){5}\pscircle(0,2){.4}
\pscircle(-2,0){.4}\pscircle(2,0){.4}
  \pscircle*[linewidth=.07](0,0){.4}
  \psline[](0,0)(0,1.7) \psline[](0,0)(1.7,0) \psline[](0,0)(-1.7,0)
   \psline[](0,0)(0,-1.7)
   \psline[linestyle=dashed](1.6,0.4)(0.4,1.6)
   \psline[linestyle=dashed](-1.6,-.4)(-0.4,-1.6)

      \psline[linestyle=dashed](1.6,-0.4)(.4,-1.6)
   \psline[linestyle=dotted](-0.4,1.6)(-1.6,.4)
   \rput(0,-2){5}\rput(0,2){3}\rput(-2,0){2}\rput(2,0){4}
\rput(1.3,1.3){$t_{Ox}$}\rput(-1.3,1.3){${t_{Ox}\over
2}$}\rput(0,-3){b)}
 }}
\end{pspicture*}

 \end{center}

 \caption{  a) The undistorted 5-atom unit with 4
flux tubes b) The distorted 5-atom unit with 1 flux tube and reduced
2-3 hopping (the barrier).
 }

\end{figure}

\vspace*{1cm}

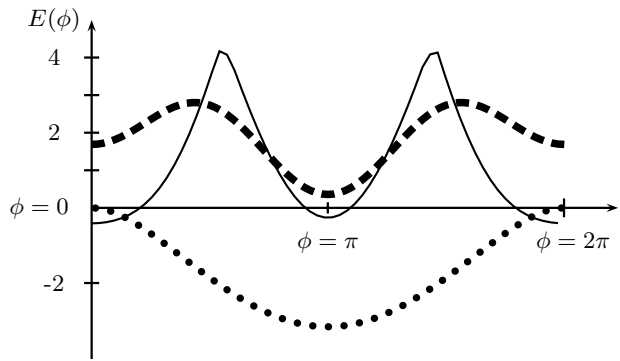
\begin{figure}
\begin{center}\psset{xunit=1cm,yunit=.5cm}
 \begin{pspicture*}(-2,-4)(8,10)
 \rput(-.5,5){$E(\phi)$}
\psline[linewidth=.03]{->}(0,0)(7,0)
\psline[linewidth=.03]{->}(0,-5)(0,5)
\rput(-.7,-.00){$\phi=0$}\psline[linewidth=.03](3.14,-.15)(3.14,.15)\rput(3.14,-.9){$\phi=\pi$}
\psline[linewidth=.03](6.28,-.25)(6.28,.25)\rput(6.4,-.9){$\phi=2\pi$}
\psline[linewidth=.03](-.1,1)(.1,1)
\psline[linewidth=.03](-.1,2)(.1,2) \rput(-.5,2){2}
\psline[linewidth=.03](-.1,3)(.1,3) 
\psline[linewidth=.03](-.1,4)(.1,4) \rput(-.5,4){4}
\psline[linewidth=.03](-.1,-2)(.1,-2) \rput(-.5,-2){-2}

 \savedata{ \uene}[
{{0, 0.}, {0.0785398, -0.00864069}, {0.157080, -0.0342294},
{0.235619,   -0.0758235}, {0.314159, -0.132021}, {0.392699,
-0.20115}, {0.471239,   -0.281448}, {0.549779, -0.371193},
{0.628319, -0.468782}, {0.706858,   -0.572773}, {0.785398,
-0.681888}, {0.863938, -0.795006}, {0.942478,   -0.911139},
{1.02102, -1.02942}, {1.09956, -1.14906}, {1.17810, -1.26938},
{1.25664, -1.38974}, {1.33518, -1.50955}, {1.41372, -1.62827},
{1.49226,   -1.7454}, {1.57080, -1.86044}, {1.64934, -1.97296},
{1.72788, -2.08251},   {1.80642, -2.18867}, {1.88496, -2.29106},
{1.96350, -2.38928}, {2.04204,   -2.48299}, {2.12058, -2.57182},
{2.19911, -2.65547}, {2.27765, -2.73361},   {2.35619, -2.80596},
{2.43473, -2.87226}, {2.51327, -2.93226}, {2.59181,   -2.98574},
{2.67035, -3.0325}, {2.74889, -3.07236}, {2.82743, -3.10519},
{2.90597, -3.13084}, {2.98451, -3.14924}, {3.06305, -3.1603},
{3.14159,   -3.16399}, {3.22013, -3.1603}, {3.29867, -3.14924},
{3.37721, -3.13084},   {3.45575, -3.10519}, {3.53429, -3.07236},
{3.61283, -3.0325}, {3.69137,   -2.98574}, {3.76991, -2.93226},
{3.84845, -2.87226}, {3.92699, -2.80596},   {4.00553, -2.73361},
{4.08407, -2.65547}, {4.16261, -2.57182}, {4.24115,   -2.48299},
{4.31969, -2.38928}, {4.39823, -2.29106}, {4.47677, -2.18867},
{4.55531, -2.08251}, {4.63385, -1.97296}, {4.71239, -1.86044},
{4.79093,   -1.7454}, {4.86947, -1.62827}, {4.94801, -1.50955},
{5.02655, -1.38974},   {5.10509, -1.26938}, {5.18363, -1.14906},
{5.26217, -1.02942}, {5.34071,   -0.911139}, {5.41925, -0.795006},
{5.49779, -0.681888}, {5.57633, -0.572773},   {5.65487, -0.468782},
{5.73341, -0.371193}, {5.81195, -0.281448}, {5.89049,   -0.20115},
{5.96903, -0.132021}, {6.04757, -0.0758235}, {6.12611, -0.0342294},
{6.20465, -0.00864069}, {6.28319, 0.}}
 ]

 \savedata{ \bicornotondo}[{{0, 1.68817}, {0.078540, 1.69665}, {0.157080, 1.72184}, {0.235619,
    1.76304}, {0.314159, 1.81908}, {0.392699, 1.88835}, {0.471239,
    1.96886}, {0.54978, 2.05825}, {0.62832, 2.15387}, {0.70686,
    2.25284}, {0.78540, 2.35209}, {0.86394, 2.44849}, {0.94248,
    2.53889}, {1.02102, 2.62021}, {1.09956, 2.68955}, {1.17810,
    2.74425}, {1.25664, 2.782}, {1.33518, 2.80087}, {1.41372,
    2.7994}, {1.49226, 2.77667}, {1.57080, 2.73226}, {1.64934,
    2.66636}, {1.72788, 2.5797}, {1.80642, 2.47355}, {1.88496,
    2.34971}, {1.96350, 2.21042}, {2.04204, 2.05833}, {2.12058,
    1.89641}, {2.19911, 1.72789}, {2.27765, 1.55616}, {2.35619,
    1.38467}, {2.43473, 1.21692}, {2.51327, 1.05628}, {2.59181,
    0.905994}, {2.67035, 0.769088}, {2.74889, 0.648298}, {2.82743,
    0.546031}, {2.90597, 0.464316}, {2.98451, 0.40477}, {3.06305,
    0.368568}, {3.14159, 0.35642}, {3.22013, 0.368568}, {3.29867,
    0.40477}, {3.37721, 0.464316}, {3.45575, 0.546031}, {3.53429,
    0.648298}, {3.61283, 0.769088}, {3.69137, 0.905994}, {3.76991,
    1.05628}, {3.84845, 1.21692}, {3.92699, 1.38467}, {4.00553,
    1.55616}, {4.08407, 1.72789}, {4.16261, 1.89641}, {4.24115,
    2.05833}, {4.31969, 2.21042}, {4.39823, 2.34971}, {4.47677,
    2.47355}, {4.55531, 2.5797}, {4.63385, 2.66636}, {4.71239,
    2.73226}, {4.79093, 2.77667}, {4.86947, 2.7994}, {4.94801,
    2.80087}, {5.0265, 2.782}, {5.1051, 2.74425}, {5.1836, 2.68955}, {5.2622,
    2.62021}, {5.3407, 2.53889}, {5.4192, 2.44849}, {5.4978,
    2.35209}, {5.5763, 2.25284}, {5.6549, 2.15387}, {5.7334,
    2.05825}, {5.8119, 1.96886}, {5.8905, 1.88835}, {5.9690,
    1.81908}, {6.0476, 1.76304}, {6.1261, 1.72184}, {6.2046,
    1.69665}, {6.2832, 1.68817}}]

 \savedata{\bicodritto}[{{0, -0.412166}, {0.1, -0.404266}, {0.2, -0.379922}, {0.3,
-0.337247}, {0.4,  -0.273239}, {0.5, -0.183993}, {0.6, -0.0649638},
{0.7, 0.0887285}, {0.8,
    0.281973}, {0.9, 0.519343}, {1., 0.804792}, {1.1, 1.14137}, {1.2,
    1.53101}, {1.3, 1.97428}, {1.4, 2.47029}, {1.5, 3.0166}, {1.6,
    3.60911}, {1.7, 4.16826}, {1.8, 4.07385}, {1.9, 3.65757}, {2.,
    3.12286}, {2.1, 2.61116}, {2.2, 2.12773}, {2.3, 1.67757}, {2.4,
    1.26534}, {2.5, 0.89535}, {2.6, 0.571445}, {2.7, 0.297022}, {2.8,
    0.0749587}, {2.9, -0.0924102}, {3., -0.203323}, {3.1, -0.256611}, {3.2,
-0.251713}, {3.3, -0.188679}, {3.4, -0.0681753}, {3.5, 0.108529},
{3.6,
    0.339575}, {3.7, 0.622534}, {3.8, 0.954439}, {3.9, 1.33182}, {4.,
    1.75073}, {4.1, 2.20682}, {4.2, 2.69536}, {4.3, 3.21131}, {4.4,
    3.74935}, {4.5, 4.08349}, {4.6, 4.13317}, {4.7, 3.50647}, {4.8,
    2.92137}, {4.9, 2.38328}, {5., 1.89601}, {5.1, 1.46174}, {5.2,
    1.08111}, {5.3, 0.753289}, {5.4, 0.476165}, {5.5, 0.246512}, {5.6,
    0.0602504}, {5.7, -0.0872677}, {5.8, -0.200953}, {5.9, -0.285651}, {6.,
-0.34582}, {6.1, -0.385232}, {6.2, -0.406711}}]



 \rput(0,0){

\dataplot[linewidth=.1,linestyle=dashed]{\bicornotondo}
 \rput(0,0){\dataplot[]{\bicodritto}
\dataplot[linewidth=.1,linestyle=dotted]{\uene}
 }

 }
\end{pspicture*}

\caption{  The ground state energy $E(\phi)$ of the CuO$_{4}$ with
$N=4$ versus flux $\phi,$ in $t_{pd}$ units;
$t_{Ox}=-0.05,\varepsilon_{p}=3.5,U=5.3$. Solid: plot of
$100*(E(\phi)-7.97)$ for the undistorted $C_{4v}$-symmetric
CuO$_{4}$ cluster with equal O-O bonds and 4 equal flux tubes.
Dashed: plot of $1000*(E(\phi)-7.97)$ for the distorted cluster.
 Notice that  we enhanced the effect 10 times
 in this case for easier comparison. Dotted:
 plot of   $50*(E(\phi)-5.126)$ for the same input data as the dashed line but
 $U=0$. In the absence of repulsion there is
 no pairing and the 4 particle system is paramagnetic.}

\end{center}

\end{figure}\label{bicornotondo}

\vspace{1cm}
\section{The one-unit model}\label{unosolo}
For simplicity, Cu$O_{4}$ clusters are natural units to build more
interesting systems.  We start from the simple 5-atom unit    and
then show how we can combine many units to build large systems in
Sect. \ref{many}. In order to have closed circuits where flux tubes
can be inserted, we consider the time-dependent Hamiltonian (see
Figure 1)

 \be H_{\rm tot} = H_{0}+H_{\tilde{t}}(t),\ee
where

  \be H_{\tilde{t}}(t)= \;
\sum_{i\sigma}\tilde{t}_{i,i+1} %
 p_{i,\sigma}^{\dagger}p_{i+1,\sigma}\;
 + h.c.
\ee

 Henceforth $t_{pd}$ will
be our energy unit; we could dispense from introducing $\e_{p},$ but
in this section  $\e_{p}=3.5$ to make the flux quantization pattern
more pronounced. There are various cases to consider.

 \subsection{Symmetric cluster}
Letting $\tilde{t}_{i,i+1}=t_{Ox}$ we have the symmetric cluster of
Figure 1 a). Moreover we insert flux tubes (see Figure 1 a)) by
setting $\tilde{t}_{i,i+1}=t_{Ox}e^{i \gamma (t)}$ with $\gamma =
\frac{2\pi \phi}{\phi_{0}} ,$ where $\phi $ is the flux in each
tube; the flux does not affect the $t_{pd}$ bonds by symmetry. We
observe a superconducting flux quantization due to the $W=0$ pairs
as shown in   Figure 2, solid line. A typical double-minimum pattern
is found with  sharp maxima due to the level crossing of paired
states with different symmetry.
 This patter is obtained with $t_{Ox}=-0.05$; if the absolute value of
the hopping $t_{Ox}$  is increased too much, the coupling to the
flux becomes excessive compared to the pairing energy and the system
turns paramagnetic\cite{topicalreview}. There are two minima at
$\phi={\phi_{0}\over 2}$ and $\phi=\phi_{0};$ in both minima, there
is pairing ($\Delta <0$). This pattern disappears together with
pairing for $U=0.$

For a time-independent flux, the $\tilde{t}$ bonds carry a screening
current, since  $\hat{j} \propto - {\de H \over \de \phi}$\cite{kohn}.  $ \phi=
\phi(t)$   is equivalent to applying an emf across each O-O bond. In
the absence of interactions, the current response to a constant emf
across e.g. the 2-3 bond is periodic due to the finite system (see
below). If the flux varies slowly, the system will follow the ground
state curve adiabatically up to the crossing point. However since
each level has conserved quantum numbers, the system cannot
adiabatically follow the ground state with slowly increasing $\phi.$
As a result, the adiabatic current is periodic with the same period
as in a normal system (e.g. the cluster with $U=0$).

\subsection{Static properties of the mildly distorted cluster}
 A more interesting behavior is obtained by breaking the square symmetry,
 as shown in Figure 1b). A
superconducting pattern is still visible (Figure 2, dashed line).
Here, $\tilde{t}_{i,i+1}=t_{ox}$ is a constant for all bonds,
except for the bond 2-3, where we take $\tilde{t}_{2,3}={t_{ox}\over
2} e^{i\g(t)}.$ No flux is present in the other O-O-Cu triangles. In
Figure 3, dashed line, we see that the superconducting double minimum
pattern is still there but the level crossing is avoided opening a
gap in the 4-fermion spectrum. However, the repulsion is needed to
produce pairing and flux quantization. The dotted line shows the
drastic effect of setting $U=0$: the system becomes paramagnetic and
normal.

\begin{figure}
\begin{center}\psset{xunit=.05cm,yunit=500cm}
 \begin{pspicture*}(-30,-.01)(300,.01)
\psline[linewidth=.03]{->}(0,0)(220,0)
\psline[linewidth=.03]{->}(0,-.005)(0,.005) \rput(-5,-.00){0}
\psline(100,-.0005)(100,.0005)\rput(100,-.002){100}
\psline(200,-.0005)(200,.0005)\rput(200,-.002){200}\rput(220,-.002){t}
\rput(10,.005){$j(t)$}
 \psline(-5,-.004)(5,-.004)\rput(-20,-.004){-.004}
  \psline(-5,.004)(5,.004)\rput(-20,.004){.004}
 \savedata{ \dynj}[{{0.,0},{2.,0.00039023},
 {4.,0.000680104},{6.,0.000992445},{8.,0.00139656},
 {10.,0.00174458},{12.,0.00200356},{14.,0.00230651},
 {16.,0.00265037},{18.,0.00288873},{20.,0.00304874},{22.,0.00323421},
 {24.,0.0033761},{26.,0.0033839},{28.,0.00331973},{30.,0.00323836},
 {32.,0.00305551},
{34.,0.00274848},{36.,0.00238365},{38.,0.00197073},
{40.,0.00146783},{42.,0.000895269},{44.,0.000294785},{46.,-0.000333476},
{48.,-0.00098466},{50.,-0.0016363},{52.,-0.00226876},{54.,-0.0028652},
{56.,-0.00341597},{58.,-0.0039138},{60.,-0.00433668},{62.,-0.00467721},
{64.,-0.00495027},{66.,-0.00513577},{68.,-0.00521335},{70.,-0.00523143},
{72.,-0.0051929},{74.,-0.00504236},{76.,-0.00482074},{78.,-0.00460815},
{80.,-0.0043272},{82.,-0.0039351},{84.,-0.00356949},{86.,-0.0032415},
{88.,-0.002801},{90.,-0.00230733},{92.,-0.00191501},{94.,-0.0014987},
{96.,-0.000956355},{98.,-0.000453925},{100.,-0.0000368729},{102.,0.000479178},
{104.,0.00105784},{106.,0.00154146},{108.,0.00201261},{110.,0.00256198},
{112.,0.00307279},{114.,0.00349916},{116.,0.00393694},{118.,0.0043509},
{120.,0.00466827},{122.,0.00492638},{124.,0.00513696},{126.,0.00524331},
{128.,0.00526329},{130.,0.00522107},{132.,0.00507089},{134.,0.00482103},
{136.,0.00452091},{138.,0.00415069},{140.,0.00368781},{142.,0.00318537},
{144.,0.002665},{146.,0.00210454},{148.,0.00152344},{150.,0.000953881},
{152.,0.000398617},{154.,-0.000130799},{156.,-0.000624265},{158.,-0.00108954},
{160.,-0.00150207},{162.,-0.00183641},{164.,-0.00213353},{166.,-0.0024056},
{168.,-0.00258765},{170.,-0.00268632},{172.,-0.00278472},{174.,-0.00285798},
{176.,-0.00281944},{178.,-0.00273193},{180.,-0.00267557},{182.,-0.00257507},
{184.,-0.00236262},{186.,-0.00214795},{188.,-0.00197187},{190.,-0.00171607},
{192.,-0.00137009},
 {194.,-0.0010591},{196.,-0.000775882},{198.,-0.000393722},{200.,0.000034182}}]
\savedata{\scostab}[
{{0,0},{2,0.000345196},{4,0.000684143},{6,0.00101073},{8,0.00131897},{10,0.00160315},
{12,0.00185783},{14,0.00207806},{16,0.00225937},{18,0.00239786},{20,0.00249025},
{22,0.00253392},{24,0.00252715},{26,0.002469},{28,0.00235939},{30,0.0021991},
{32,0.00198978},{34,0.00173412},{36,0.0014356},{38,0.00109856},{40,0.000728059},
{42,0.000329841},{44,-0.0000896802},{46,-0.000523687},{48,-0.000965033},
{50,-0.00140638},{52,-0.00184032},{54,-0.00225956},{56,-0.00265706},
{58,-0.00302618},{60,-0.00336075},{62,-0.00365516},{64,-0.00390463},
{66,-0.0041051},{68,-0.00425331},{70,-0.0043469},{72,-0.00438428},
{74,-0.00436487},{76,-0.0042889},{78,-0.00415738},{80,-0.00397213},
{82,-0.00373562},{84,-0.00345112},{86,-0.00312242},{88,-0.00275387},
{90,-0.00235029},{92,-0.00191688},{94,-0.00145922},{96,-0.000983152},
{98,-0.000494692},{100,0},{102,0.000494692},{104,0.000983152},{106,0.00145922},
{108,0.00191688},{110,0.00235029},{112,0.00275387},{114,0.00312242},{116,0.00345112},
{118,0.00373562},{120,0.00397213},{122,0.00415738},{124,0.0042889},{126,0.00436487},
{128,0.00438428},{130,0.0043469},{132,0.00425331},{134,0.0041051},{136,0.00390463},
{138,0.00365516},{140,0.00336075},{142,0.00302618},{144,0.00265706},{146,0.00225956},
{148,0.00184032},{150,0.00140638},{152,0.000965033},{154,0.000523687},
{156,0.0000896802},{158,-0.000329841},{160,-0.000728059},{162,-0.00109856},
{164,-0.0014356},{166,-0.00173412},{168,-0.00198978},{170,-0.0021991},
{172,-0.00235939},{174,-0.002469},{176,-0.00252715},{178,-0.00253392},
{180,-0.00249025},{182,-0.00239786},{184,-0.00225937},{186,-0.00207806},
{188,-0.00185783},{190,-0.00160315},{192,-0.00131897},{194,-0.00101073},
{196,-0.000684143},{198,-0.000345196},{200,0}}]




 \rput(0,0){
\dataplot{\dynj} \dataplot[linestyle=dashed]{\scostab}

 }

\end{pspicture*}

\caption{ The dynamical  current response in units of $\frac{e
t_{pd}}{\hbar}$ (solid line) versus time flowing in the 2-3 bond
with $\phi(t)=\h t,$ $\h={\pi\over 100}.$ This value of $\h$ is
below the  gap in the spectrum, which is of order $|t_{Ox}|=0.05$. }

\end{center}

\label{scossadyn}
\end{figure}
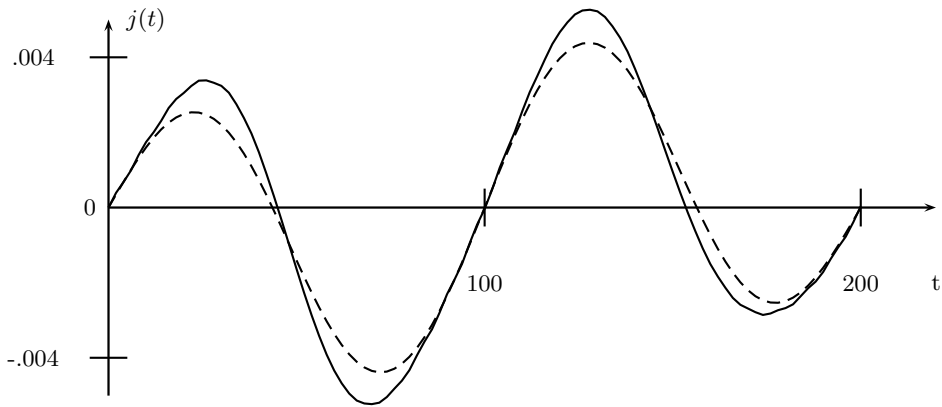

\vspace{1cm}
\subsection{Emf and frequency of currents}
 In the paired situation  of Figure 2, dashed line,
  we will consider   the time-dependent
case, $\phi/\phi_{0} \rightarrow Vt$, where $V$ plays the role of an
electromotive force, and $t$ in  units of $\frac{\hbar}{t_{pd}}$. An
adiabatic behavior is now achieved when the time dependence is slow
compared to the gap. We are using the Gauss system, hence  \be
\phi_{0}=\frac{h c}{e}\ee and the Faraday law reads \be
V=-\frac{1}{c}\frac{\de \phi}{\de t}.\ee

Therefore, we introduce the frequency $\h$ by setting \be \h
=-\frac{e V}{\hbar}.\ee

Equation (\ref{Josephson}) becomes \be \frac{I}{I_{J}}=\sin(2\h
t)\ee for superconductor junctions, while the 2 is missing for
normal systems.

This estimate agrees with the results of Figure 3. The solid line is
the current through the 2-3 bond obtained by integrating the
Schr\"odinger equation. The initial conditions are $\phi=0$ with the
system in the ground state with $N$=4. The dotted line, reported for
comparison, is the adiabatic response $j_{ad}(t)=c\frac{\de
E(4)}{\de \phi(t)}.$ The two curves agree fairly well.  Since
$t_{Ox}\ll 1,$ the main contribution to the current response comes
from the 123 plaquette, which responds with frequency $\h$ for
normal systems and $2\h$ for superconducting behavior. Some current
contribution involves other plaquettes as well, perhaps mixing
different frequencies. The agreement with the adiabatic
approximation confirms the present analysis and will be used in
Sect.IV . The normal response would be periodic with half the
frequency.

\begin{figure}
\begin{center}\psset{xunit=.05cm,yunit=50cm}
 \begin{pspicture*}(-30,-.1)(300,.1)

\psline[linewidth=.03]{->}(0,0)(220,0)
\psline[linewidth=.03]{->}(0,-.09)(0,.09) \rput(-5,-.00){0}
\psline(100,-.005)(100,.005)\rput(100,.0155){100}
\psline(200,-.005)(200,.005)\rput(200,-.015){200}\rput(220,-.002){t}
\rput(20,.05){$j(t)$}
 \psline(-5,-.05)(5,-.05)\rput(-20,-.05){-.05}
  \psline(-5,.05)(5,.05)\rput(-20,.05){.05}
 \savedata{ \dynj}[{{0., 0}, {2., -0.00339753}, {4., -0.00691894}, {6., -0.0103962},
{8.,   -0.0137411}, {10., -0.0171257}, {12., -0.0205779}, {14.,
-0.0239042}, {16.,   -0.0271086}, {18., -0.0303438}, {20.,
-0.0335343}, {22., -0.0365201}, {24.,   -0.0393614}, {26.,
-0.042131}, {28., -0.0446921}, {30., -0.0469448}, {32., -0.0489428},
{34., -0.0506688}, {36., -0.0519953}, {38., -0.0528676}, {40.,
-0.0532935}, {42., -0.0532319}, {44., -0.0526219}, {46.,-0.0514222},
{48.,   -0.0496312}, {50., -0.0472926}, {52., -0.0444153}, {54.,
-0.0409684}, {56.,   -0.0370504}, {58., -0.0328378}, {60.,
-0.0283548}, {62., -0.0236344}, {64., -0.0189498}, {66.,
-0.0145099}, {68., -0.0102919}, {70., -0.00645534}, {72.,
-0.00334274}, {74., -0.000995043}, {76., 0.000660623}, {78.,
    0.00138565}, {80.,
    0.0010407}, {82., -0.000137375}, {84., -0.00208146}, {86., -0.0049123},
{88., -0.00837457}, {90., -0.012096}, {92., -0.0160462}, {94.,
-0.0200579},   {96., -0.023637}, {98., -0.0265761}, {100.,
-0.0288549}, {102., -0.0301104},   {104., -0.0300441}, {106.,
-0.0287835}, {108., -0.0263032}, {110.,   -0.022374}, {112.,
-0.0171807}, {114., -0.0110685}, {116., -0.00404562},   {118.,
0.00375753}, {120., 0.0118364}, {122., 0.0199012}, {124.,
    0.0278897}, {126., 0.0354504}, {128., 0.0421876}, {130.,
    0.0480498}, {132., 0.0529852}, {134., 0.0567652}, {136.,
    0.0593559}, {138., 0.0609155}, {140., 0.0614922}, {142.,
    0.0611443}, {144., 0.0600891}, {146., 0.0585274}, {148.,
    0.0566046}, {150., 0.0545216}, {152., 0.0524576}, {154.,
    0.0505249}, {156., 0.0488609}, {158., 0.0475654}, {160.,
    0.0466234}, {162., 0.0460493}, {164., 0.0458844}, {166., 0.046033}, {168.,
     0.0463657}, {170., 0.0468801}, {172., 0.0475216}, {174.,
    0.0480965}, {176., 0.0485386}, {178., 0.0488999}, {180., 0.049077}, {182.,
     0.0489324}, {184., 0.0485548}, {186., 0.0480335}, {188.,
    0.0472482}, {190., 0.0461872}, {192., 0.0450442}, {194.,
    0.0438533}, {196., 0.0424827}, {198., 0.0410148}, {200., 0.0396242}}
] \savedata{\scostab}[{{0, 0}, {2, -0.00703588}, {4, -0.0138933},
{6, -0.0204247}, {8, -0.0265082},  {10, -0.0320627}, {12,
-0.0370449}, {14, -0.0414509}, {16, -0.0452972}, {18,   -0.0486157},
{20, -0.0514473}, {22, -0.0538359}, {24, -0.0558241}, {26,
-0.0574515}, {28, -0.0587536}, {30, -0.0597617}, {32, -0.060503},
{34,   -0.0610001}, {36, -0.0612724}, {38, -0.0613363}, {40,
-0.0612056}, {42,   -0.0608921}, {44, -0.0604054}, {46, -0.0597542},
{48, -0.0589457}, {50, -0.0579865}, {52, -0.0568822}, {54,
-0.0556383}, {56, -0.0542595}, {58,   -0.0527507}, {60, -0.0511165},
{62, -0.0493612}, {64, -0.0474896}, {66,   -0.0455063}, {68,
-0.043416}, {70, -0.0412236}, {72, -0.0389342}, {74,   -0.036553},
{76, -0.0340857}, {78, -0.0315377}, {80, -0.0289151}, {82,
-0.0262238}, {84, -0.0234702}, {86, -0.0206607}, {88, -0.0178019},
{90,   -0.0149005}, {92, -0.0119635}, {94, -0.00899782}, {96,
-0.00601057}, {98, -0.0030089}, {100, 0}, {102, 0.0030089}, {104,
0.00601057}, {106,
    0.00899782}, {108, 0.0119635}, {110, 0.0149005}, {112, 0.0178019}, {114,
    0.0206607}, {116, 0.0234702}, {118, 0.0262238}, {120, 0.0289151}, {122,
    0.0315377}, {124, 0.0340857}, {126, 0.036553}, {128, 0.0389342}, {130,
    0.0412236}, {132, 0.043416}, {134, 0.0455063}, {136, 0.0474896}, {138,
    0.0493612}, {140, 0.0511165}, {142, 0.0527507}, {144, 0.0542595}, {146,
    0.0556383}, {148, 0.0568822}, {150, 0.0579865}, {152, 0.0589457}, {154,
    0.0597542}, {156, 0.0604054}, {158, 0.0608921}, {160, 0.0612056}, {162,
    0.0613363}, {164, 0.0612724}, {166, 0.0610001}, {168, 0.060503}, {170,
    0.0597617}, {172, 0.0587536}, {174, 0.0574515}, {176, 0.0558241}, {178,
    0.0538359}, {180, 0.0514473}, {182, 0.0486157}, {184, 0.0452972}, {186,
    0.0414509}, {188, 0.0370449}, {190, 0.0320627}, {192, 0.0265082}, {194,
    0.0204247}, {196, 0.0138933}, {198, 0.00703588}, {200, 0}}
]




 \rput(0,0){
\dataplot{\dynj} \dataplot[linestyle=dashed]{\scostab}

 }

\end{pspicture*}

\caption{  The dynamical  current response in units of $\frac{e
t_{pd}}{\hbar}$ (solid  line) versus time flowing in the 2-3 bond
with $\phi(t)=\h t,$ $\h={\pi\over 100},$  as above, but $U=0$.}

\end{center}

\label{scossadynuzero}
\end{figure}
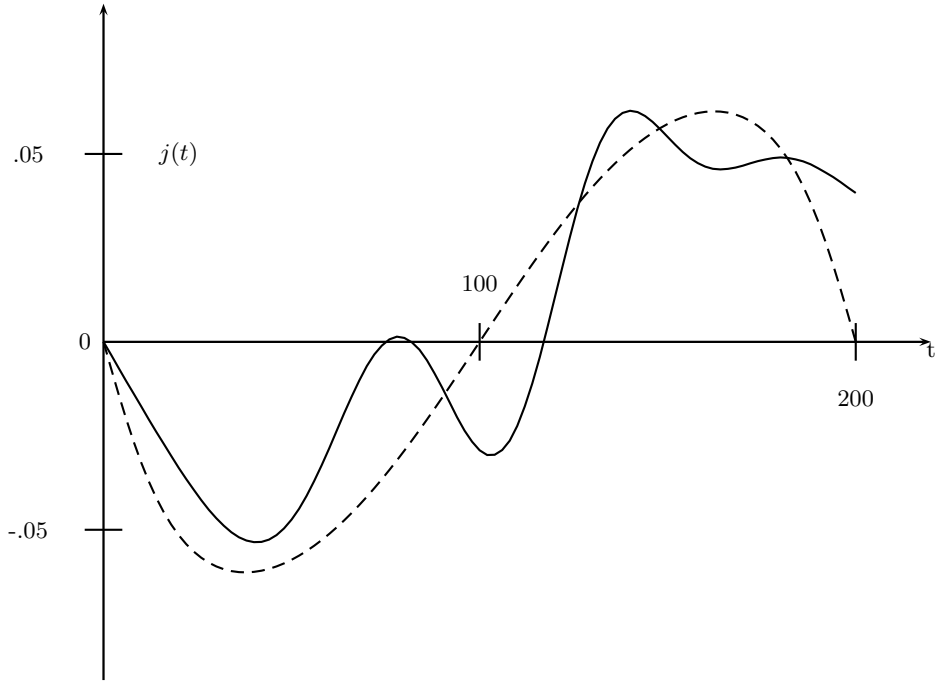

   The system becomes
normal for $U=0$, a fact which is apparent from the magnetic
response (dotted line of Figure 2). The current response through the
2-3 bond also become normal, as we can see    from the halved
frequency of the current response  in Figure 4. Note that the
adiabatic result (dashed) does not agree very well with the response
obtained by integrating the Schr\"odinger equation (solid). In fact,
the gap $\sim t_{Ox}$ is missing in the spectrum.
\subsection{Supercurrent flowing without bias}
In a macroscopic junction one can excite a supercurrent that will
last forever if the exciting bias is removed, unless there are
dissipative elements elsewhere in the circuit. This is also borne
out in the paired situation  of Figure 2, dashed line;
  we will consider   the time-dependent
case, $\phi/\phi_{0} \rightarrow Vt$, where $V$ plays the role of an
emf of finite duration.
 In Figure 5 we plot
the current flowing  across the 2-3 bond (barrier) when the emf
lasts from $t=0$ to $t=20$. At $t=0$, $\phi={\pi\over 2}.$ A
constant supercurrent is superimposed on the oscillatory response.

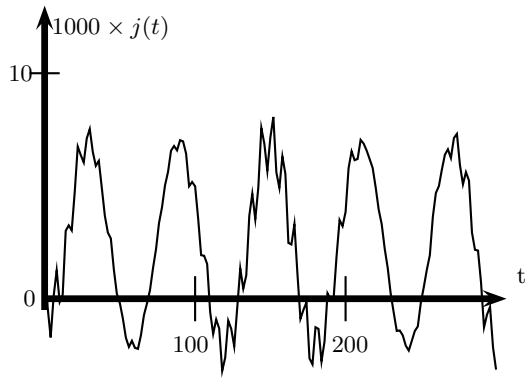
\begin{figure}\label{supercurrent}
\begin{center}\psset{xunit=.02cm,yunit=.3cm}
 \begin{pspicture*}(-50,-10)(350,14)

 \savedata{ \sc}[{{0., 1.95852}, {2., -0.388431}, {4., -1.71563}, {6., 0.125295}, {8.,
    1.25514}, {10., -0.142065}, {12., 0.18156}, {14., 3.01152}, {16.,
    3.25397}, {18., 3.02989}, {20., 4.73774}, {22., 6.74612}, {24.,
    6.37004}, {26., 6.04349}, {28., 7.11287}, {30., 7.52644}, {32.,
    6.51148}, {34., 5.89423}, {36., 6.11418}, {38., 4.9115}, {40.,
    3.68518}, {42., 2.93595}, {44., 2.67111}, {46., 1.28883}, {48.,
    0.2147}, {50., -0.198674}, {52., -0.884409}, {54., -1.73262}, {56.,
-2.13665}, {58., -1.86547}, {60., -2.17729}, {62., -2.22425}, {64.,
  -1.61898}, {66., -0.725377}, {68., -0.211513}, {70., 0.379398},
{72.,
    1.3457}, {74., 2.18897}, {76., 3.35281}, {78., 4.05684}, {80.,
    5.02922}, {82., 5.62354}, {84., 6.56703}, {86., 6.773}, {88.,
    6.57912}, {90., 7.03085}, {92., 6.97787}, {94., 6.43454}, {96.,
    4.99046}, {98., 5.17798}, {100., 4.98676}, {102., 3.57104}, {104.,
    1.93182}, {106., 1.90384}, {108.,
    1.53892}, {110., -0.591887}, {112., -1.56174}, {114., -0.921184}, {116.,
-1.31967}, {118., -3.19637}, {120., -2.61113}, {122., -0.972093},
{124.,   -1.33233}, {126., -2.1136}, {128., -0.396576}, {130.,
1.34509}, {132.,
    0.44853}, {134., 1.03629}, {136., 3.675}, {138., 4.65431}, {140.,
    3.52416}, {142., 4.93016}, {144., 7.57571}, {146., 6.87993}, {148.,
    5.67181}, {150., 7.14217}, {152., 8.06102}, {154., 5.60166}, {156.,
    4.91767}, {158., 6.30941}, {160., 5.53436}, {162., 2.4753}, {164.,
    2.40496}, {166., 3.30637}, {168.,
    1.10372}, {170., -0.889454}, {172., -0.492567}, {174., -0.15945}, {176.,
-2.64472}, {178., -2.90844}, {180., -1.26387}, {182., -1.33809},
{184.,   -2.77665}, {186., -1.89845}, {188.,
    0.240428}, {190., -0.0161561}, {192., -0.0169497}, {194., 1.87445}, {196.,
     3.49134}, {198., 3.21707}, {200., 3.84957}, {202., 5.8004}, {204.,
    6.53657}, {206., 6.14536}, {208., 6.19765}, {210., 7.04679}, {212.,
    6.88114}, {214., 6.57191}, {216., 6.1913}, {218., 5.79964}, {220.,
    4.93594}, {222., 3.95426}, {224., 3.30906}, {226., 2.21913}, {228.,
    1.36571}, {230.,
    0.213513}, {232., -0.620098}, {234., -1.42815}, {236., -1.40884}, {238.,
-1.62454}, {240., -2.08277}, {242., -2.29763}, {244., -1.87148},
{246.,   -1.4329}, {248., -1.22776}, {250., -0.179705}, {252.,
0.524209}, {254.,
    1.28533}, {256., 1.93232}, {258., 3.63636}, {260., 4.689}, {262.,
    4.9923}, {264., 5.72825}, {266., 6.38641}, {268., 6.62747}, {270.,
    6.24138}, {272., 7.12521}, {274., 7.31121}, {276., 5.92716}, {278.,
    5.07528}, {280., 5.61388}, {282., 5.24234}, {284., 2.93274}, {286.,
    2.16897}, {288., 2.13624}, {290.,
    0.732647}, {292., -1.2715}, {294., -0.704986}, {296., -0.360477}, {298.,
-2.16418}, {300., -3.14361}}]

 \rput(0,0){\dataplot[]{\sc}}
\psline[linewidth=.09]{->}(0,0)(307,0)\rput(317,1){t}\rput(-10,0){0}
\psline[linewidth=.09]{->}(0,-.5)(0,13)\rput(44,12){$1000\times
j(t)$}
 \psline(100,-1)(100,1)\rput(97,-2){100}
 \psline(200,-1)(200,1)\rput(203,-2){200}
 \psline(-10,10)(10,10)\rput(-16,10){10}\psline(-10,20)(10,20)
\end{pspicture*}

\caption{ Current response in units of $\frac{e t_{pd}}{\hbar}$ of
the mildly distorted CuO$_{4}$ unit. A constant  emf across the 2-3
bond (barrier)  lasts from $t=0$ to $t=20$. At $t=0$,  The current
response to a flux $\phi(t)=\gamma_{0}+\eta t $ where
$\eta={\pi\over 100}$ as above and $\gamma_{0}={\pi\over 2}.$   A
constant supercurrent is superimposed to  the oscillations.}

\end{center}

\end{figure}

\subsection{Shapiro-like  effect}
Still in the paired situation  of Figure 2, dashed line,
  we also  consider   the time-dependent
case  of an oscillating field.  In superconducting junctions there
are spikes in the current-voltage characteristics when the voltage
has a constant component and an oscillating one at radio-frequency
$\w_{r}$ of amplitude $V_{r}$ (Shapiro effect): \be
V_{tot}=V+V_{r}\cos(\omega_{r}t).\ee Then Eq.(\ref{Josephson})
reads:   \be I(t)=I_{J}\sin\left[ \frac{2eV t}{\hbar
}+\gamma_{0}+\frac{2eV_{r}}{\hbar\omega_{r}}\sin(\omega_{r}t)
\right].\ee In particular one finds a zero-voltage spike, because at
$V=0$ the resulting $I(t)$ has a DC component. We can mimic such a
behavior with our model. In Figure 6 we plot the current response
through the 2-3 bond  to a flux \be
\phi(t)=\g_{0}+2\frac{\h_{r}}{\w_{r}}\sin(\w_{r}t)\ee with
$\g_{0}={\pi \over 2},$  $\h_{r}=\frac{\pi}{200}$ and $\w_{r}={\pi
\over 10}.$ We observe the typical Shapiro-like DC current with a
ripple on it.

\begin{figure}
\begin{center}\psset{xunit=.02cm,yunit=.3cm}
 \begin{pspicture*}(-50,-2)(350,34)

 \savedata{ \deputato}[{{0., 14.0638}, {2., 13.3481}, {4., 13.2668}, {6., 13.8482}, {8.,
    14.9202}, {10., 16.4677}, {12., 18.0533}, {14., 18.8785}, {16.,
    18.7535}, {18., 18.0056}, {20., 16.7589}, {22., 15.1407}, {24.,
    13.852}, {26., 13.4889}, {28., 13.8068}, {30., 14.2868}, {32.,
    14.8019}, {34., 15.0752}, {36., 14.5483}, {38., 13.1979}, {40.,
    11.6519}, {42., 10.2931}, {44., 9.26125}, {46., 9.05635}, {48.,
    9.99221}, {50., 11.4755}, {52., 12.7524}, {54., 13.7404}, {56.,
    14.3678}, {58., 14.2144}, {60., 13.4318}, {62., 12.863}, {64.,
    12.8535}, {66., 13.3179}, {68., 14.4575}, {70., 16.2635}, {72.,
    17.8884}, {74., 18.6474}, {76., 18.7027}, {78., 18.2049}, {80.,
    16.9283}, {82., 15.3218}, {84., 14.2943}, {86., 14.0338}, {88.,
    14.1979}, {90., 14.7243}, {92., 15.4309}, {94., 15.6221}, {96.,
    14.8913}, {98., 13.609}, {100., 12.1262}, {102., 10.5286}, {104.,
    9.34167}, {106., 9.21585}, {108., 10.05}, {110., 11.2217}, {112.,
    12.4238}, {114., 13.4922}, {116., 13.9716}, {118., 13.6247}, {120.,
    12.9413}, {122., 12.4529}, {124., 12.3138}, {126., 12.756}, {128.,
    14.0895}, {130., 15.9519}, {132., 17.4894}, {134., 18.3613}, {136.,
    18.6608}, {138., 18.2102}, {140., 16.9478}, {142., 15.5437}, {144.,
    14.6838}, {146., 14.4014}, {148., 14.5706}, {150., 15.2571}, {152.,
    16.0151}, {154., 16.083}, {156., 15.3315}, {158., 14.1409}, {160.,
    12.5785}, {162., 10.8115}, {164., 9.57881}, {166., 9.45438}, {168.,
    10.1052}, {170., 11.0704}, {172., 12.2682}, {174., 13.3239}, {176.,
    13.5986}, {178., 13.1383}, {180., 12.5308}, {182., 12.0154}, {184.,
    11.7162}, {186., 12.1795}, {188., 13.6869}, {190., 15.5209}, {192.,
    16.9927}, {194., 18.0649}, {196., 18.584}, {198., 18.0938}, {200.,
    16.8788}, {202., 15.7372}, {204., 14.985}, {206., 14.5874}, {208.,
    14.8786}, {210., 15.8092}, {212., 16.5273}, {214., 16.4732}, {216.,
    15.8806}, {218., 14.7844}, {220., 13.0116}, {222., 11.127}, {224.,
    10.0106}, {226., 9.76992}, {228., 10.1067}, {230., 11.0412}, {232.,
    12.3131}, {234., 13.1638}, {236., 13.2178}, {238., 12.8322}, {240.,
    12.2287}, {242., 11.4751}, {244., 11.1104}, {246., 11.7579}, {248.,
    13.2246}, {250., 14.9163}, {252., 16.5163}, {254., 17.8054}, {256.,
    18.3024}, {258., 17.8139}, {260., 16.8331}, {262., 15.8471}, {264.,
    15.0338}, {266., 14.7302}, {268., 15.2747}, {270., 16.2616}, {272.,
    16.9158}, {274., 16.952}, {276., 16.491}, {278., 15.3105}, {280.,
    13.4282}, {282., 11.5928}, {284., 10.4746}, {286., 10.0355}, {288.,
    10.2647}, {290., 11.2268}, {292., 12.4265}, {294., 13.0494}, {296.,
    13.0042}, {298., 12.6243}, {300., 11.8837}}]

 \rput(0,0){\dataplot[]{\deputato}}
\psline[linewidth=.09]{->}(0,0)(307,0)\rput(317,1){$t$}
\psline[linewidth=.09]{->}(0,-.5)(0,23)\rput(44,22){$1000\times
j(t)$}
 \psline(100,-1)(100,1)\rput(100,-1){100}
 \psline(200,-1)(200,1)\rput(200,-1){200}
 \psline(-10,10)(10,10)\rput(-16,10){10}\psline(-10,20)(10,20)\rput(-16,20){$20$}
\end{pspicture*}
\vspace{1cm}

\caption{ The  current response in units of $\frac{e t_{pd}}{\hbar}$
to a flux $\phi(t)=\gamma_{0}+{2\h\over \omega_{r}
}\sin(\omega_{r}t)$ where $\gamma_{0}={\pi\over 2}$ and
$\omega_{r}={\pi\over 10}.$  The Shapiro-like constant component is
superimposed to a ripple due to the alternative paths available to
the current.}

\end{center}

\end{figure}
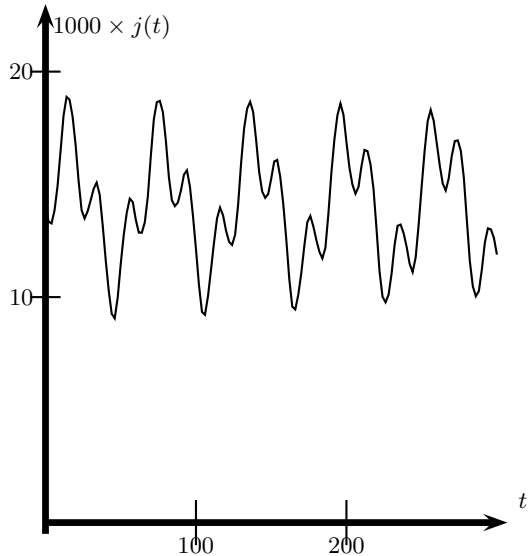\label{shapiro}

\vspace{1cm}

\subsection{Comment on the one-unit model}
It appears from the above that the 5-atom cluster with suitable
parametrization can simulate all the characteristic signatures of
Josephson junctions, as if it already contained a seed of the order
parameter. This is indeed surprising, but we show below that
combining the small units we can go smoothly to large systems still
keeping the same characteristic behavior; simplicity stems from the
fact that if there are many pairs they move coherently.

\section{The many-units model}\label{many}

In This Section we study the time-dependent propagation of many $W=0$ pairs along rings
of CuO$_{4}$ Hubbard clusters in presence of a barrier.

Our system is made of $L-1$ CuO$_{4}$ units plus a "barrier"
standing between the 1-st and the $L-1$-th units. The barrier itself
is taken to be a CuO$_{4}$ cluster, but with very high on-site
energies, in order to make its occupancy only virtual.

The $L-1$ units are connected each other by the usual inter-cluster
hopping Hamiltonian $H_{\tau}$, with real hopping integral $\tau$,
while the hopping between the barrier and the 1-st and the $L-1$-th
units is assumed to be complex and time-dependent:  $\tau ' e^{i
\gamma (t)}$, where the phase $\gamma (t)$ will be discussed later.
By means of such a choice we have in mind that in transport
gedankenexperiments the potential drops only through the barrier.

According to the notation of our previous work\cite{sfq}, the model reads: \be
H_{\rm tot} = H_{0}+H_{\tau}+H_{\rm barrier}+H_{\tau '}(t)\ee
 with
$$ H_{0}=\sum_{\alpha =1}^{L-1} [ t_{pd} \sum_{i\sigma}( d^{\dagger}_{\alpha \sigma}p_{\alpha, i\sigma}+
p_{\alpha, i\sigma}^{\dagger}d_{\alpha \sigma})+ U(n^{(d)}_{\alpha
\uparrow} n^{(d)}_{\alpha \downarrow}+\sum_{i}n^{(p)}_{\alpha, i
\uparrow}n^{(p)}_{\alpha, i\downarrow})], $$ where,
$p^{\dagger}_{\alpha, i\sigma}$ is the  creation operator for a hole
onto the Oxygen $i=1,..,4$ of the $\alpha$-th cell and so on. In
this section $\e_{p}=0.$

$H_{\tau}$ is an inter-cell hopping Hamiltonian which allows a fermion
in the $i$-th Oxygen site of the $\alpha$-th unit to move towards
the  $i$-th Oxygen site of the $\alpha \pm 1$-th unit (without
involving the $L$-th unit, which is the barrier) with hopping
integral $\tau$:
$$
H_{\tau}=\sum_{\alpha=1}^{L-2}\sum_{i\sigma} \tau\;
p_{\alpha,i\sigma}^{\dagger}p_{\alpha +1,i\sigma}\; + h.c. .
$$
Finally we model the barrier with the following Hamiltonian: \be
H_{\rm barrier}=t_{pd} \sum_{i\sigma}( d^{\dagger}_{L \sigma}p_{L,
i\sigma}+ p_{L, i\sigma}^{\dagger}d_{L \sigma})+ U(n^{(d)}_{L
\uparrow} n^{(d)}_{L \downarrow}+\sum_{i}n^{(p)}_{L, i
\uparrow}n^{(p)}_{L, i\downarrow}) + WN_{tot}, \ee where $N_{tot}$
is the number of particles in the $L$-th cluster and $W$ is an extra
(large) energy felt by each particle in the barrier, which can be
provided, for instance, by a gate voltage. The barrier is linked to
the rest of the ring by a complex hopping Hamiltonian: \be
H_{\tau'}(t)=\tau' e^{i \gamma (t)} \; \sum_{i\sigma}
 (p_{L-1,i\sigma}^{\dagger}p_{L,i\sigma}\; + p_{L,i\sigma}^{\dagger}p_{1,i\sigma})\; + h.c. \equiv
\tau' e^{i \gamma (t)} H_{J} +h.c. \ee

In the static case we set (with the same notation of \cite{sfq}) \be
\gamma = \frac{2\pi \phi}{ \Lambda \phi_{0}} ,\ee where $\phi$ is
the total flux through the ring and $\Lambda=2$ as long as the flux
is associated with two bonds. In the time-dependent case, we will
set $\phi/\phi_{0} \rightarrow Vt$, where $V$ plays the role of an
electromotive force, as in the previous Section.

\subsection{Static Case}

Let us begin with the static case. As usual we introduce
$H_{\tau}+H_{\tau'}$ perturbatively, in the same spirit of our
previuos work\cite{sfq}. When $U/t\simeq 5$, $\tau=\tau'=0$ and the number of particles is
appropriate and even, each CuO$_{4}$ cluster is populated by 2 or 4 particles
(since the occupations with 3 particles have a gap of the order of
$|\Delta|$ ), while the barrier is totally empty, since the extra
energy
 even for a single occupancy is very large.
Now, if $|\tau|,|\tau'|<<|\Delta|$, the perturbations are operative
only by the second order. In particular $H_{\tau}$ provides an
effective hopping $\tau_{eff}\sim \tau^{2}/|\Delta|$ for bound $W=0$
pairs, which can propagate in the 1,2,...,$L-1$
 clusters as hard-core bosons.
On the other hand $H_{\tau'}$  acts differently. Indeed the
occupancy of barrier is highly unfavoured in the ground state
because of the large energy $W$. As a consequence the $W=0$ pair can
jump from the 1-st cluster to the $L-1$-th cluster an viceversa only
at the fourth order in $H_{\tau'}$. In this case the effective
hopping through the barrier is $ J e^{i  \frac{4\pi \phi}{ \phi_{0}}
} $ with \be J= -\frac{1}{2}(\tau')^{4}\frac{1}{|\Delta|
\varepsilon^{2}}C , \ee where \be \varepsilon = E(3)+E(2)+E(1)+W \ee
(with $E(N)$ the ground state energy of the CuO$_{4}$ with $N$
particles), and C is a order 1 costant given by

 \begin{equation}
 \begin{array}{c}
C=\langle \Psi_{L-1}(4)| \otimes \langle \Psi_{L}(0) |\otimes
\langle \Psi_{1}(2) | \ H_{J}  |
  \Psi_{1}(2)   \rangle \otimes | \Psi_{L}(1)  \rangle\otimes |\Psi_{L-1}(3)
   \rangle \times
\\
\langle \Psi_{L-1}(3)| \otimes \langle \Psi_{L}(1) |\otimes \langle
\Psi_{1}(2) | \ H_{J}  |
  \Psi_{1}(3)   \rangle \otimes | \Psi_{L}(0)  \rangle\otimes |\Psi_{1}(3)  \rangle \times
 \\
\langle \Psi_{L-1}(3)| \otimes \langle \Psi_{L}(0) |\otimes \langle
\Psi_{1}(3) | \ H_{J}  |
  \Psi_{1}(3)   \rangle \otimes | \Psi_{L}(1)  \rangle\otimes |\Psi_{L-1}(2)  \rangle \times
 \\
\langle \Psi_{L-1}(2)| \otimes \langle \Psi_{L}(1) |\otimes \langle
\Psi_{1}(3) | \ H_{J}  |
  \Psi_{1}(2)   \rangle \otimes | \Psi_{L}(0)  \rangle\otimes |\Psi_{1}(4)  \rangle ,
\end{array}
\end{equation}
 where $|\Psi_{\alpha}(N)  \rangle$ is the ground state of the
$\alpha$-th cluster with $N$ particles.

The resulting low-energy effective hamiltonian is then equivalent to
a 1-dimensional spin $1/2$ chain with ${\cal L} =L-1$  sites, given
by (see Ref. \cite{bic2} for the case of $\tau' =0$): \be
H_{eff}(\phi)=-2\tau_{eff} \sum_{\alpha=1}^{{\cal L}-1}
S^{z}_{\alpha}S^{z}_{\alpha + 1}+ \tau_{eff} \sum_{\alpha=1}^{{\cal
L}-1}  (S^{+}_{\alpha}S^{-}_{\alpha + 1}+h.c.)
-\tau_{eff}S^{z}_{{\cal L}}S^{z}_{ 1}+ (J e^{i  \frac{4\pi \phi}{
\phi_{0}} }S^{+}_{{\cal L}}S^{-}_{1}+h.c.). \ee

We observe that the barrier has been "integrated out" and the
resulting model has the 1-st and the ${\cal L}$-th sites linked by
$J$.

It is worth to note that this model, because of the the presence of
two different parametres $\tau_{eff}$ and $J$, is not solvable by
the Bethe ansatz, therefore we must resort numerical methods. We
solved $H_{eff}(\phi)$ with ${\cal L}$ up to 12 and $S^{z}_{tot}$ up
to 3. In this case, in the original fermionic model there were
$L=13$ CuO$_{4}$ clusters (12 + barrier) and $6$ $W=0$ pairs, that is $12
\times 2 + 6\times 2=36$ fermionic particles. In other words the
number of $W=0$ pairs is $2 \times S^{z}_{tot}$, which is a
conserved quantity.

In figure we show the ground state energy $E(\phi)$ as a function of
the magnetic flux for  ${\cal L}=12$ and $2\times
S^{z}_{tot}=1,2,3,4,5,6$. It is remarkable that whatever is the
number of $W=0$ pairs in the original systems (1,2,3,4,5,6), the
peridicity of $E(\phi)$ does not change and shows the same
superconducting flux quantization. This is a very interesting
feature, suggesting that all the $W=0$ pairs may behave
"coherently", providing the same macroscopic response, no matter of
their number. It is also clear that the oscillations of $E(\phi)$
get more and more pronounced as the number of pairs in the ground
state become of the same order of number of the lattice sites. This
is an indication that such a behavior should survive in the
thermodynamic limit.

\begin{figure}
\includegraphics[width =7.5cm ]{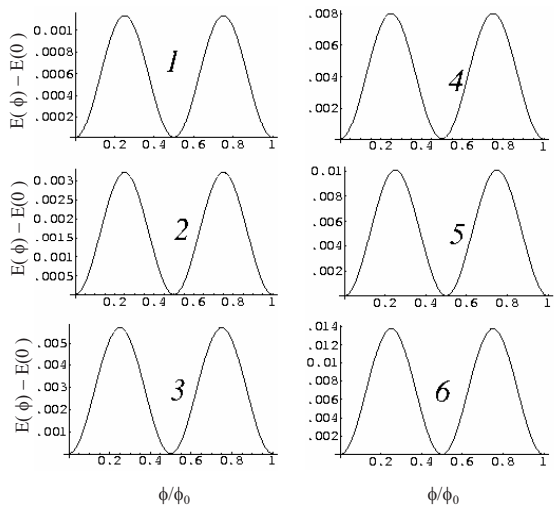}
\caption{$E(\phi)-E(0)$ (in units of $t_{pd}$) as a function of $\phi/\phi_{0}$  for $2 \times S^{z}_{tot}=1,2,3,4,5,6$. Here we have
taken and $J/\tau=0.1$.} \label{}
\end{figure}

Finally we observe that the trend of $E(\phi)$ by incresing the
number of particles is deeply different with respect to the one
observed in the repulsive Hubbard ring. In the latter case\cite{yufowler}
\cite{kusmartsev}, there are many minima between $\phi=0$
and $\phi=\phi_{0}$, depending on the ratio $Nt/LU$, where $N$ is
the number of particles and $L$ is the number of lattice sites. On
the other hand, in the case under consideration it seems that the
superconducting nature of the charge carriers "freezes" the period
oscillations to be $\phi_{0}/2$.

As a last remark, we notice that $E(\phi)$ is a smooth function of
$\phi$. This means that no level crossing occurs between $\phi=0$
and $\phi=\phi_{0}$. This is a direct consequence of the breaking of
the translational symmetry induced by the presence of the barrier.
As discussed in the previous Section, this condition will be important in the time-dependent case, where, at least
in the adiabatic approximation, the system at $t=0$ (supposed to be
in the ground state at $\phi=0$) will be able to evolve continuosly
to the static ground state at $\phi=\phi_{0}/2$. This condition is
necessary in order to have a time-dependent superconducting
response.

\subsection{Time-Dependent Case}

Now we consider a time-dependent perturbation, with the effective
hopping through the barrier  given by
$
J e^{i  4\pi Vt }.
$
In this case we rely on the adiabatic approximation, such that,
nevertheless of the time-dependence, the low energy Hamiltonian is
the same as $H_{eff}$, with $\phi/\phi_{0} \rightarrow Vt$. The
correctness of this approximation has been checked directly in the previous Section,
in the time-dependent analysis of the single CuO$_{4}$ cluster. There, the
exact time-dependent solution of the whole original electronic
problem was achieved and compared to the approximated one.

The exact time-dependent solution of the Schr\"odinger equation
associated to $H_{eff}(t)$ is much harder than the static solution.
Therefore we only consider the case ${\cal L}=4$ and $2\times
S^{z}_{tot}=2$. Anyway, in the original fermionic model, it means
$L=5$ CuO$_{4}$ clusters (4+barrier) and 2 $W=0$ pairs, that is $12$
particles. As initial condition at $t=0$ we take the gruond state of
$H_{eff}(\phi)$ at $\phi =0$.

Our aim is to evaluate the time-dependent current $I(t)$ through the
barrier, i.e. through the bond governed by $J e^{i  \frac{4\pi Vt}{
\phi_{0}} }$, according to \be I(t)=J \langle \Psi(t) | e^{i  4\pi
Vt }S^{+}_{{\cal L}}S^{-}_{1}- e^{-i  4\pi Vt }S^{-}_{{\cal
L}}S^{+}_{1} | \Psi(t) \rangle. \ee

The exact numerical solution shows that $I(t)$ is a peridic function
of $t$ with period $1/(2V)$. In figure we plot $I(t)$ as a function
of $Vt$ in two different unitary intervals: (0,1) and (48,49).

\begin{figure}
\includegraphics[width =4.5cm ]{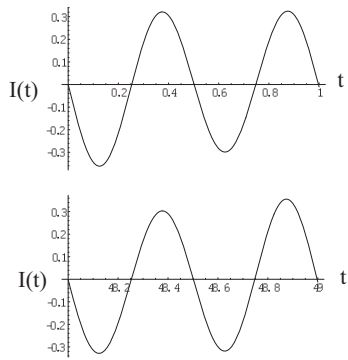}
\caption{$I(t)$ as a function of $Vt$ for $2 \times  S^{z}_{tot}=2$
for two intervals (0,1) and (49,40):  Here we have taken and
$J/\tau=0.1$.} \label{}
\end{figure}

We stress that the time-dependent response of the many-units model have the same Josephson-like behaviour of the single-unit model.

\section{Conclusions}\label{Conclusion}

We have shown how to model a Josephson-like junction using
microscopic ingredients like electrons hopping in  an atomic
lattice, a tunneling barrier, and on-site correlation. The barrier
consists of a bond characterized by a weak time-dependent  hopping,
where the emf is applied. The current response is easier to
calculate and analyze when there is emf on a single bond. We stress
that the Hubbard interaction $U$ leads to a current response in
quantitative agreement with Eq. (\ref{Josephson}). This happens in
circumstances that we understand, namely: i) the formation of $W=0$
pairs, ii)  a well developed double minimum in the energy-flux plot,
iii) a soft distortion of the symmetry opening a small gap (due to
avoided crossing) in the many-body spectrum, iv) a small enough emf
$V$, in order to allow for adiabatic response. When the last
condition is realized,
 calculations based on the adiabatic approximation are in good agreement
with the full solutions; this allows us to extend the analysis
to fairly large systems. In modeling rings of Hubbard clusters, the
barrier was formed by a CuO$_{4}$ unit with high one-electron
energy; in order to deal with the low-energy physics, we derived an effective
 Hamiltonian where the barrier is integrated out and represented by
 an effective, weakened, complex bond, similar to the barrier in
 the above distorted CuO$_{4}$ case.

In the single-unit model, the superconducting  AC current has the
same frequency ${2e V\over \hbar}$ as one would expect to observe in
a macroscopic junction; in the latter case, however, the normal
response is  a DC current. In our cluster calculations, on the other
hand, the current response of the normal systems is periodic with
half the Josephson frequency; this happens with one-particle systems
and with 4-body ones with $U=0$, when no pairing occurs.  This is
due to a size effect that we wish to comment upon here.

 As shown in Ref. (\onlinecite{cini80}), in the non-interacting
case (which is clearly normal) the transient current across a tunnel
barrier between infinite leads tends to a constant for long times;
  the asymptotic $t\rightarrow \infty$  behavior yields the current-voltage
characteristics calculated by several people by time-independent
approaches \cite{landauer}\cite{caroli}\cite{feuchtwang}\cite{meir}.
This has recently been proven as a theorem by Stefanucci and
Almbladh\cite{stefanucci04}  within the Time-Dependent Density
Functional Theory in the Local Density Approximation. In order to understand
the\index{Stefanucci} difference, we recall that the infinite
junctions considered in (\cite{cini80}) never reach equilibrium, and
the steady current that holds in the $t\ra \inf$ limit is just a
never-ending transient. On the contrary, the oscillating behavior of
the small cluster follows almost adiabatically the instantaneous
ground state. Another difference is that the normal system is unable
to trap the magnetic flux: when the field penetrates the sample our
modeling should be less appropriate than in superconductors. So we
can expect that the normal current oscillation at half frequency,
that we observe in the small cluster, should acquire a longer and
longer transient behaviour with increasing sample size, while the
field penetration should superimpose different oscillation
frequencies. Instead, the superconducting pairs should keep their
frequency in line with equation (\ref{Josephson}) right to the
thermodynamic limit.

 According to our  interpretation, the Josephson
effect is essentially another facet of the superconducting flux
quantization. The necessary distortion is always present in
macroscopic systems, where in addition a large energy separates the
states of different pair symmetry. Unbiased supercurrents and
Shapiro spikes  are also obtained by the same models that mimic the
AC Josephson behaviour.

In the many-units model,  the CuO$_{4}$  clusters are  combined in
large systems linked by weak hopping integrals;   the solution is
found by mapping the problem into a spin 1/2 chain.  In the
time-independent case, we presented evidence that the flux
quantization pattern is unmodified when several bound pairs
propagate together, except that the energy barriers separating the
minima grow large. In the time-dependent case we solved
Schr\"odinger equation numerically, and again we got a
Josephson-junction-like behavior.

 Finally, it is remarkable  that even with one pair,
  that can be on the left
or on the right, there is enough uncertainty on the pair number to
create some ghost order parameter.

\section*{Acknowledgements}
E. P. was supported by INFN grant 10068.


\end{document}